\documentstyle[12pt]{article}

\begin{document}

\baselineskip=7mm
\renewcommand{\arraystretch}{1.3}

\newcommand{\cf}{{ f}}
\newcommand{\TeV}{\,{\rm TeV}}
\newcommand{\GeV}{\,{\rm GeV}}
\newcommand{\MeV}{\,{\rm MeV}}
\newcommand{\keV}{\,{\rm keV}}
\newcommand{\eV}{\,{\rm eV}}
\newcommand{\Tr}{{\rm Tr}\!}
\newcommand{\be}{\begin{equation}}
\newcommand{\ee}{\end{equation}}
\newcommand{\bea}{\begin{eqnarray}}
\newcommand{\eea}{\end{eqnarray}}
\newcommand{\ba}{\begin{array}}
\newcommand{\ea}{\end{array}}
\newcommand{\bmat}{\left(\ba}
\newcommand{\emat}{\ea\right)}
\newcommand{\refs}[1]{(\ref{#1})}
\newcommand{\ler}{\stackrel{\scriptstyle <}{\scriptstyle\sim}}
\newcommand{\ger}{\stackrel{\scriptstyle >}{\scriptstyle\sim}}
\newcommand{\lag}{\langle}
\newcommand{\rag}{\rangle}
\newcommand{\ns}{\normalsize}

\begin{titlepage}
\title{{\Large\bf Supersymmetry Hierarchy Problems and Anomalous 
Horizontal U(1) Symmetry }\\
                          \vspace{-4.5cm}
                          \hfill{\ns KAIST-TH 15/96\\}
                          \hfill{\ns CBNU-TH 961107\\}
                          \hfill{\ns hep-ph/9611293\\}
                          \vspace{3.5cm} }
\author{Kiwoon Choi${}^\dagger$,\hspace{.2cm}  
        Eung Jin Chun${}^*$,\hspace{.2cm}  and \hspace{.2cm}  
        Hyungdo Kim${}^\dagger$\\[.5cm]
  {\ns\it Department of Physics, Korea Advanced Institute of Science
           and Technology}\\
  {\ns\it Taejon 305-701,  Korea${}^\dagger$} \\[.5cm]
  {\ns\it Department of Physics, Chungbuk National University}\\
  {\ns\it Cheongju, Chungbuk 360-763,  Korea${}^*$} \\ 
        }
\date{}
\maketitle
\begin{abstract} \baselineskip=7.2mm {\ns
It is suggested that various hierarchy problems in supersymmetric
standard model, i.e. the Yukawa hierarchies, the $\mu$ problem,
and the suppression of dangerous baryon and/or lepton number
(B/L) violating couplings,  
are resolved altogether in the framework of 
horizontal U(1) symmetry whose spontaneous breaking results 
in the appearance of one expansion parameter (the Cabibbo angle).
Within a reasonable range of U(1) charges, there exist a few models 
compatible with experiments.  The specific sizes of   
B/L violating couplings of these models are calculated and 
several phenomenological consequences are discussed.
\\[.3cm]
PACS numbers: 11.30.Hv, 12.15.Ff, 12.60.Jv
}\end{abstract}

\thispagestyle{empty}
\end{titlepage}

\section*{\ns\bf 1. Introduction}

In some sense, the minimal supersymmetric standard model (MSSM) suffers 
from more
hierarchy problems than the standard model (SM). The gauge invariance under 
SU(3)$_c \times$ SU(2)$_L \times$ U(1)$_Y$ would allow the
following superpotential,
\bea \label{mssm}
  W_{\rm MSSM} &&= \mu_0 H_1 H_2 + Y^u_{ij} H_2 Q_i U^c_j + 
         Y^d_{ij} H_1 Q_i D^c_j + Y^e_{ij} H_1 L_i E^c_j \nonumber\\
  &&+\mu_i L_i H_2 + \Lambda^u_{ijk}  U^c_i D^c_j D^c_k + 
         \Lambda^d_{ijk} L_i Q_j D^c_k + \Lambda^e_{ijk} L_i L_j E^c_k
\nonumber \\ 
  &&+{\Gamma_{ijk}^l \over M_P} Q_i Q_j Q_k L_l  +
 {\Gamma_{ijk}^0 \over M_P} Q_i Q_j Q_k H_1  +
 {\Gamma_{ijk}^{'l} \over M_P} U^c_i U^c_j D^c_k E^c_l+..., 
\eea
where $Y$'s and $\Lambda$'s are Yukawa couplings, $\mu_{0}$ and $\mu_i$ are
dimension-one parameters, $\Gamma$'s denote the coefficients
of B/L violating $d=5$ operators, and the other big letters denote the 
superfields of Higgses, quarks and leptons.  
Concerning the above superpotential, 
one fundamental question which applies also for the SM
is why the quark and lepton masses 
are hierarchical,
e.g. why the up quark Yukawa coupling $Y_u\simeq 10^{-5}$
is much smaller than the top quark coupling $Y_t \simeq 1$.
Unlike the case of the SM, the baryon number (B) and the lepton number (L)
violating Yukawa couplings ($\Lambda$'s) generate also a kind of 
hierarchy problem since they  are required to be highly  suppressed.
For instance, proton stability 
forces $\Lambda^u$ and/or $\Lambda^d$ to be extremely small: 
$\Lambda^u \Lambda^d \leq 10^{-24}$ \cite{proton}.  
Another hierarchy problem concerns the mass parameters, $\mu_0$ and $\mu_i$.
The Higgs mass parameter $\mu_0$ should be of order of  the
electroweak scale.  The $\mu$ problem \cite{kn}
consists in understanding why 
$\mu_0$ is so small  compared to the fundamental
scale of the theory, e.g.  the Planck mass $M_P$:  
$\mu_0/M_P \simeq
10^{-16}$.  The parameters $\mu_i$ are required to be
further suppressed by the
smallness of neutrino masses \cite{hs} unless one assumes a special
form of soft supersymmetry breaking \cite{hemp}--\cite{nar}.  
Finally even the coefficients of 
B/L violating $d=5$ operators,  i.e. $\Gamma$'s, are required
to be suppressed to a certain degree, e.g.  $\Gamma^i_{112}\leq 10^{-8}$
and $\Gamma^0_{12j}\Lambda^d_{ijk}\leq 10^{-8}$.

It is certainly appealing to assume that the above-mentioned hierarchies 
in $W_{\rm MSSM}$ have a common origin.  
Recently, the pattern of quark mass matrices 
are studied  in the framework of supergravity (SUGRA)  model 
\cite{lns}--\cite{cl}
in which nonrenormalizable couplings of quarks and leptons to a SM singlet 
field $\phi$ are constrained by a horizontal abelian symmetry U(1)$_X$ 
to generate Yukawa hierarchies \cite{fn}.  
The vacuum expectation value of a singlet $\phi$ which breaks
U(1)$_X$ yields the expansion parameter of Yukawa couplings:
$$
\lambda=\lag \phi \rag/M_P\simeq 0.22 \quad ({\rm Cabibbo \,\, angle}).
$$
It has been noted   that the $\mu$ problem can be resolved
also by means of U(1)$_X$ \cite{nir}.
In this scheme,  supersymmetry breaking is assumed to occur
spontaneously in a hidden sector and
is transmitted to the
observable sector by supergravity interactions.  
The size of supersymmetry breaking in the observable sector
is of order 
$m_{3/2}$ which can be identified as the electroweak scale.
Then for a certain U(1)$_X$ charge assignment \cite{nir},
$\mu$ appears to be of order  $\lambda m_{3/2}$ as a consequence
of the U(1)$_X$ selection rule.
As was discussed recently, 
the horizontal symmetry U(1)$_X$  can be useful also for
suppressing   the dangerous  B/L violating
couplings \cite{blr,cl}.

An interesting feature of the model with U(1)$_X$ 
is its connection to superstring
theory.  In the simple model with one expansion parameter $\lambda=\langle
\phi\rangle/M_P$, the
observed quark mass eigenvalues requires the Green-Schwarz mechanism
to cancel the anomalies \cite{gs}.  The ratio between the
anomalies would be determined by the canonical value of $\sin^2{\theta_w} =
3/8$ at the string scale \cite{iba}.
In this paper, we  show how
a horizontal abelian gauge symmetry 
compatible with the observed quark masses and mixing 
can  constrain the B/L violating operators and also  the 
$\mu$ terms to be phenomenolgically safe.
We then pick out several viable models with a reasonable
range of U(1)$_X$ charges and discuss their 
phenomenological consequences.

In the models we found, all hierarchies
in the MSSM superpotential, i.e. the hierarchical fermion masses and mixings,
the  hierarchically small $\mu$,  and finally the hierarchically small 
B/L violating couplings including those of nonrenormalizable terms,
can be understood by  the U(1)$_X$ selection rule alone.
It turns out that there exists only one such model (Model 1) 
if the maximum magnitude of the  U(1)$_X$ charges is limited 
to be less than 10
for the basic unit of charge normalized to one.
There appear several more models (Models 2 and 3 for instance)
if one relaxes the limit to 15.
Although quite attractive in the sense that all hierachies
have a common origin, 
we feel that the models, particularly Models 2 and 3,  have a flaw that
the magnitudes of the required  U(1)$_X$ charges are
still  big (although not unreasonably big)
in view of the 
the anomalous U(1) charges in various string model
constructions \cite{charges}.
This would make their appearance as a low energy limit
of string theory
not very plausible.
In this regard, an interesting possibility  is that the model
contains another spontaneously broken gauge
symmetry (in addition to U(1)$_X$) 
which would be responsible
for the weak scale value of $\mu$ and/or  the suppression
of some B/L violating couplings \cite{choi}.
The U(1)$_X$ charges in this context  can be smaller and thus
fit better for string theory.

\bigskip

\section*{\ns\bf 2. Basic properties}

The quark mixing matrix $V_{\rm CKM}$ in the Wolfenstein parameterization
\cite{wol} is approximately given by
\be \label{ckm}
 V_{\rm CKM} \simeq  \left( \ba{ccc} 
       1 & \lambda & \lambda^3 \\
       \lambda & 1 & \lambda^2 \\
       \lambda^3 & \lambda^2 & 1 \ea\right)
\ee
where all the coefficients of order 1 are omitted.
The class of models under consideration assume that the Cabibbo angle 
originates from the spontaneous breaking of U(1)$_X$ as
$\lambda = \lag \phi \rag/M_P$.
Under the additional  assumption that U(1)$_X$ breaking
is described entirely  by the order parameter $\lambda$,
the eigenvalues of up and down quark 
masses at the Planck scale  are given by \cite{rrr}:
\bea \label{qeigen}
 \left( M^u \right)_{\rm diagonal} \simeq m_t (\lambda^8, \lambda^4, 1) 
        \,,\nonumber \\
 \left( M^d \right)_{\rm diagonal} \simeq m_b (\lambda^4, \lambda^2, 1)
       \,.
\eea
As shown in ref.~\cite{cl}, two informations in eqs.~\refs{ckm} and
\refs{qeigen} are enough to reconstruct the corresponding up and down
quark mass matrices in our scheme.  The observed up and down quark 
masses and mixing determine the six U(1)$_X$ charges of the MSSM
superfields.
Throughout this paper, we will use the small letters 
$q,u,d,l,e,h_1,h_2$ to denote the U(1)$_X$
charges of the corresponding MSSM superfields.  The charge of $\phi$
can be any integer, say $-N$. But for the purpose of convenience we
will normalize it to $-1$ which means that the charges of the 
MSSM superfields can be fractional numbers with $N$ in the
denominator. Note that this means that 
the  MSSM possesses an unbroken $Z_N$ parity for $N\geq 2$.
The large top quark mass says that the top quark 
Yukawa coupling comes from the renormalizable term 
$H_2 Q_3 U^c_3$ in the SUGRA superpotential,  and thus
$$
 h_2+q_3+u_3=0.
$$
The bottom quark
Yukawa coupling could well be obtained by the nonrenormalizable term 
$H_1Q_3D^c_3(\phi/M_P)^x$ where 
the  positive integer $x$ is given by
$$
x = h_1+q_3+d_3.
$$
Here $x$ can be $0,1,2$ or $3$
with  $\tan\beta \simeq \lambda^x m_t/m_b$.
Denoting $q_{ij} \equiv q_i-q_j$ {\it etc.}, the charge assignments
\bea \label{qch}
  {\rm (I)}& &(q_{13},q_{23})=(3,2),\quad (u_{13},u_{23}) = (5,2),\quad
  (d_{13},d_{23}) = (1,0) \nonumber \\
  {\rm (II)}& &(q_{13},q_{23})=(-3,2),\quad (u_{13},u_{23}) = (11,2),\quad
  (d_{13},d_{23}) = (7,0) 
\eea
are known to yield the acceptable quark Yukawa matrices \cite{blr}.
However, as we will see, the pattern (II) does not yield 
any acceptable model for the range of U(1)$_X$ charges 
not exceeding  15 when  the basic unit of charge is normalized to
unity.
Therefore here we quote the up and down quark 
Yukawa matrices only for the pattern (I):
\be \label{qyuk}
 Y^u \simeq \bmat{ccc} \lambda^8 & \lambda^5 & \lambda^3 \\
                       \lambda^7 & \lambda^4 & \lambda^2 \\
                       \lambda^5 & \lambda^2 & 1  \emat  \,,\quad
 Y^d \simeq \lambda^x \bmat{ccc} \lambda^4 & \lambda^3 & \lambda^3 \\
                       \lambda^3 & \lambda^2 & \lambda^2 \\
                       \lambda & 1 & 1  \emat \,.
\ee

\bigskip
As a gauge symmetry, the horizontal symmetry U(1)$_X$
has to be anomaly free.
The mixed anomalies of SU(3)$_c^2$--U(1)$_X$, 
SU(2)$_L^2$--U(1)$_X$, U(1)$_Y^2$--U(1)$_X$ and U(1)$_Y$--U(1)$_X^2$ are
given by  
\bea
 A_3 &=& \sum_i(2 q_i + u_i + d_i), \nonumber\\
 A_2 &=& \sum_i (3 q_i + l_i) + (h_1+h_2),  \\
 A_1 &=& \sum_i ({1\over3}q_i+{8\over3}u_i+{2\over3}d_i+l_i+2 e_i)+
           (h_1+h_2), \nonumber\\
 A'_1&=& \sum_i(q_i^2-2u_i^2+d_i^2-l_i^2+e_i^2)-(h_1^2-h_2^2)\nonumber\,.
\eea
As shown by Binetruy-Ramond \cite{br}, 
the observed quark masses are not compatible with 
the usual anomaly-free condition; $A_3=A_2=A_1=0$.
But the MSSM with  U(1)$_X$ symmetry may come from superstring 
theory which allows the Green-Schwarz mechanism of anomaly cancellation
\cite{gs}. Furthermore, the gauge coupling 
unification near the Planck scale can be understood in terms of 
the Green-Schwarz mechanism when the anomalies satisfy the relation;
$A_3:A_2:A_1=1:1:5/3$ \cite{iba}.  Therefore we assume that the horizontal 
symmetry U(1)$_X$ is a gauge symmetry coming from  superstring theory.
In this case, the identity $A_1+A_2-8 A_3/3=0$ implies
\be \label{iden}
 h_1+h_2 = \sum_i(q_{i3}+d_{i3}) - \sum_i(l_{i3}+e_{i3}) \,.
\ee
Throughout this paper, we assume that $h_1+h_2=-1$ for which
$\mu_0$ appears to be of order
the weak scale as a consequence of U(1)$_X$. (See the subsequent discussion
on the $\mu$ parameters.)
Then combined with the $b$--$\tau$ unification condition,
$$
l_3+e_3=q_3+d_3,
$$
the above relation from the anomaly cancellation  provides an information
on the charged lepton Yukawa couplings $Y^e$. 
Since $\sum_i(q_{i3}+d_{i3})=6$ from eq.~\refs{qch}, we have
$\sum_i(l_{i3}+e_{i3})=7$, implying ${\rm det}(Y^e)=\lambda^{3x+7}$.
Then the observed charged lepton masses indicate that the
eigenvalues of $Y^e$ are given by
\be \label{eeigen}
 \left( Y^e \right)_{\rm diagonal} = \lambda^x (\lambda^{5},
          \lambda^{2}, 1) \,.
\ee
It is also useful to recall that the desired charged lepton mass 
matrix with the above eigenvalues follows from the charge relations
\cite{cl}
\be\label{ech}
(e_{13}, e_{23})=(5-l_{13}, 2-l_{23})\,, \quad 
(e_{13}, e_{23})=(9-l_{13}, -2-l_{23}). 
\ee

In order to discuss how the couplings other than $Y^{u,d,e}$ and also
the $\mu$ terms 
are constrained by the spontaneously broken U(1)$_X$, 
one needs to write down the most general K\"ahler
potential invariant under
SU(3)$_c\times$ SU(2)$_L\times$ U(1)$_Y\times$ U(1)$_X$ gauge symmetry.
The U(1)$_X$ distinguishes  the Higgs doublet $H_1$ from the lepton
doublet $L_i$. (We call $H_1$ the field having the largest
$\mu$, viz $\mu_0 \geq \mu_i$.)
Let us write down the K\"ahler potential containing 
only $L_i$ and $H_{1,2}$;
\bea \label{K}
K&& = \frac{1}{2}H_1 H_1^\dagger + \frac{1}{2}H_2 H_2^\dagger  +
   \frac{1}{2}L_i L_j^\dagger \left[\left(\phi\over M_P\right)^{l_i-l_j}\theta(l_i-l_j)
     + \left(\phi^\dagger\over M_P\right)^{l_j-l_i}\theta(l_j-l_i)
       \right] \nonumber\\
 && +L_i H_1^\dagger \left[\left(\phi\over M_P\right)^{l_i-h_1}\theta(l_i-h_1)
        + \left(\phi^\dagger\over M_P\right)^{h_1-l_i}\theta(h_1-l_i) \right]
        \\
 && + H_1 H_2 \left( \phi^\dagger\over M_P\right)^{-h_1-h_2}\theta(-h_1-h_2)
  + L_i H_2 \left( \phi^\dagger\over M_P\right)^{-l_i-h_2}\theta(-l_i-h_2)
     +{\rm h.c}, \nonumber
\eea
where $\theta(y) =1$ if $y$ is a {\it non-negative integer}, 
and $\theta(y)=0$ otherwise.
Note that the holomorphic operators $H_1 H_2$ and $L_i H_2$ can
appear in the K\"ahler potential as well as in the superpotential.

The ``effective'' MSSM superpotential \refs{mssm} generated after the 
spontaneous breaking of both supersymmetry and U(1)$_X$ contains the
$\mu$ terms given by 
\bea \label{mus}
  \mu_0 &=& M_P \lambda^{h_1+h_2}_\theta +
          m_{3/2} \bar{\lambda}^{-h_1-h_2}_\theta \,,  \nonumber\\
  \mu_i &=& M_P \lambda^{l_i+h_2}_\theta +
          m_{3/2} \bar{\lambda}^{-l_i-h_2}_\theta  \,,
\eea
where  the first terms in the right hand sides arise  from
the underlying SUGRA superpotential, while the second terms
are the contributions
from the SUGRA K\"{a}hler potential.
Here $\lambda^x_\theta \equiv \lambda^x\theta(x)$ 
and $\bar{\lambda} =\lambda^*$.
For the desirable value of $\mu_0 \simeq m_{3/2}$, the charge
$h_1+h_2$ may happen to be $h_1+h_2 = 23\sim 25$, yielding
$\mu_0=M_P\lambda^{h_1+h_2}\simeq m_{3/2}$.
In our approach, however, the anomaly-free condition
does not allow such a large value of $h_1+h_2$
[see eq.~\refs{iden}].
As noted by Nir \cite{nir}, the acceptable fermion mass matrices
are compatible with the choice $h_1+h_2=-1$, which may actually solve the
``$\mu_0$ problem'' in the context of horizontal symmetry.
Therefore, in this paper we will assume 
\be
h_1+h_2=-1,
\ee
for which  $\mu_0\simeq
\bar{\lambda}m_{3/2}$.
The smallness of $\mu_i$ can be understood in the similar manner.
Even though the anomaly-free condition allows a large positive
value of $l_i+h_2$,   one may still assume that $l_i+h_2$ are
all negative, and thus 
$\mu_i\simeq m_{3/2}\bar{\lambda}^{|l_i+h_2|}$.

Although we assume $h_1+h_2=-1$ in this paper, another
choice of $h_1+h_2=0$ can also
give rise to acceptable fermion mass matrices while satisfying the 
anomaly-free condition \refs{iden}. However then  
we need an
independent  mechanism, e.g. other spontaneously broken
gauge symmetry \cite{choi}, ensuring  $\mu_0$ to be of order the weak
scale since the horizontal
symmetry allows $\mu_0$ to be of order $M_P$.

The Yukawa couplings of  other renormalizable operators appearing
in the effective MSSM superpotential are given by 
\bea \label{couplings}
 Y^d_{ij} &=& \lambda^{h_1+q_i+d_j}_\theta, \quad
 Y^e_{ij} \, = \, \lambda^{h_1+l_i+e_j}_\theta, \nonumber\\
 \Lambda^d_{ijk} &=& \lambda^{l_i+q_j+d_k}_\theta + 
      {m_{3/2}\over M_P} \bar{\lambda}^{-l_i-q_j-d_k}_\theta\,,\nonumber\\
 \Lambda^e_{ijk} &=& \lambda^{l_i+l_j+e_k}_\theta + 
      {m_{3/2}\over M_P} \bar{\lambda}^{-l_i-l_j-e_k}_\theta\,,\\
 \Lambda^u_{ijk} &=& \lambda^{u_i+d_j+d_k}_\theta + 
      {m_{3/2}\over M_P} \bar{\lambda}^{-u_i-d_j-d_k}_\theta \,. \nonumber
\eea
where we ignored the 
the  K\"ahler potential contributions, i.e. 
the parts suppressed by
$m_{3/2}/M_P$, to $Y^{d,e}$.
Note that the up and down quark Yukawa couplings 
$Y^{u,d}$ are already given in
eq.~\refs{qyuk}.  

The above $\mu$'s in eq.~\refs{mus} and the Yukawa couplings in 
eq.~\refs{couplings}
are given in the non-canonical basis where the
K\"ahler metric has off-diagonal components:
\bea \label{kahler}
 K_{Q_i Q^\dagger_j}= \lambda_\theta^{|q_{ij}|}\,, &
 K_{U^c_i U^{c\dagger}_j}= \lambda_\theta^{|u_{ij}|}\,, &
 K_{D^c_i D^{c\dagger}_j}= \lambda_\theta^{|d_{ij}|}\,,  \nonumber\\
 K_{L_i L^\dagger_j}= \lambda_\theta^{|l_{ij}|}\,, &
 K_{E^c_i E^{c\dagger}_j}= \lambda_\theta^{|e_{ij}|}\,, &
 K_{L_i H^\dagger_1}= \lambda_\theta^{|l_i-h_1|}\,.
\eea
The above K\"{a}hler metric can be diagonalized
as $(K)_{\rm diagonal}=
(U^{\dagger}KU)$ where $U$ 
takes the same form as the K\"ahler metric in the order of
magnitude estimate \cite{dps,blr}.
This diagonalization  would  alter the original estimate of  
the $\mu$'s and the couplings 
in eqs.~\refs{mus} and \refs{couplings}.
Especially, 
diagonalizing away the K\"{a}hler metric components
$K_{L_i H_1^\dagger}$  leads to the change
\bea \label{diago}
 \mu_i &\longrightarrow& \lambda_\theta^{|l_i-h_1|}\mu_0 + \sum_j
        \lambda_\theta^{|l_{ij}|}\mu_j \,, \\
 \Lambda^d_{ijk} &\longrightarrow& 
       \sum_{n,p}\lambda_\theta^{|l_i-h_1|+ |q_{jn}|+ |d_{kp}|} Y^d_{np} + 
       \sum_{m,n,p}\lambda_\theta^{|l_{im}|+ |q_{jn}|+ |d_{kp}|} 
                 \Lambda^d_{mnp} \,. \nonumber
\eea
As we will see, the above change of $\Lambda^d$ is essential 
for  constraining the charges $l_i$ from the
experimental bounds on $\Lambda^d$ since it is related to
$Y^d$ which is known to us as eq~\refs{qyuk}.
However the change of $\mu_i$
is not so relevant for us since 
it does not change the size of $\mu_i$ for the models under consideration.

We have  to yet consider two more redefinitions of the couplings 
which would alter the estimated size of the couplings.
First, normally the Yukawa couplings $\Lambda^d$ are defined  
after the $\mu_i$ terms, i.e.  $\mu_iL_i H_2$, in the superpotential
are rotated away.
This results in an additional  contribution to
$\Lambda^d$, which is  given by $\delta\Lambda_{ijk}
\simeq Y^d_{jk} \mu_i/\mu_0$.
In some cases, e.g. the case (iii) of the section 3, 
this contribution  becomes dominant and thus alters
the order of magnitude estimate of
$\Lambda^d$, while in other cases, e.g. the case (ii) of the section (3),
it doesn't.

Another possibility is the change of couplings in the course
of going to the mass eigenstates in order to make
a contact with experiments. 
The quark and lepton Yukawa matrices
$Y^I$ ($I=u,d,e$) can be diagonalized  by 
biunitary transformations 
\be \nonumber
    \left( Y^I \right)_{\rm diagonal} = U_I Y^I V_I^\dagger\,.
\ee
For us, the unitary matrix $U_I$ can be decomposed into three
rotations described by the small  angles $S^I_{12}, S^I_{13}$ and $S^I_{23}$; 
\be \label{Us}
 U_I \simeq \bmat{ccc} 1 & -S^I_{12} & 0 \\ 
                          S^I_{12} & 1 & 0 \\
                          0 & 0 & 1 \emat
          \bmat{ccc} 1 & 0 & -S^I_{13}  \\ 
                          0 & 1 & 0 \\
                          S^I_{13}  & 0 & 1 \emat
          \bmat{ccc} 1 & 0 & 0  \\ 
                          0 & 1 & -S^I_{23} \\
                          0 & S^I_{23}  & 1 \emat \,,
\ee
also similarly for $V_I$ with $S^{\prime I}_{12}, S^{\prime I}_{13}$ 
and $S^{\prime I}_{23}$ \cite{hr}. The general expressions for $S^I_{ij}$ 
and $S^{\prime I}_{ij}$ are calculated in refs.~\cite{dps,blr}.
For the acceptable quark Yukawa matrices given by the charge
assignments (I) and (II) of eq.~\refs{qch},
it is easy to find that 
\bea \label{angles}
 & S^I_{12} \simeq Y^I_{12}/Y^I_{22} \,,\quad
   S^I_{13} \simeq Y^I_{13} \,,\quad
   S^I_{23} \simeq Y^I_{23} \,, &  \nonumber \\
 & S^{\prime I}_{12} \simeq Y^I_{21}/Y^I_{22} \,,\quad
   S^{\prime I}_{13} \simeq Y^I_{31} \,,\quad
   S^{\prime I}_{23} \simeq Y^I_{23} \,, & 
\eea
where $I=u,d$.  
In fact, we find also  that 
the above expressions of the rotation angles are applicable also
for the lepton Yukawa matrices satisfying all the
phenomenological constraints.  
Therefore, the expressions of eq.~\refs{angles} are valid for all
$I=u,d,e$ in our scheme.
For the biunitary transformations   
defined by the angles of eq.~\refs{angles},
the diagonalization of the quark and lepton mass matrices  
gives the same effect on
the order of magnitudes of the couplings as
the diagonalization of the K\"{a}hler metric.
As a consequence, the mass diagonalization does not change further the order
of magnitudes of the couplings once the effects of
the K\"{a}hler metric diagonalization are taken into
account as  eq.~\refs{diago}.

\bigskip

Combining all the U(1)$_X$ charge relations discussed so far
with the anomaly-free conditions
$$
A_3=A_2=3A_1/5, \quad  A'_1=0, 
$$
we are left with 
four independent charges, for instance $l_i$ and $x$.
In section 4, 
we vary $l_i$ and $x$ under the condition
that $x=0,1,2,$ or 3 to find some reasonable 
charge assignments which fulfill the bounds on B/L violating 
couplings which will be discussed in section 3.

\bigskip

\section*{\ns\bf 3. Constraints on B/L violating terms}

Let us first discuss in detail the L violating terms.
The existence of $\mu_i$ or $\Lambda^{d,e}$ plays an 
important role of generating significant neutrino masses 
when their values are not too small \cite{hs}.
Therefore, it is interesting to see whether some phenomenologically
observable neutrino masses and mixing \cite{smi}
can arise naturally in our scheme.
As seen from eqs.~\refs{couplings} and \refs{diago}, $\mu_i$
and  $\Lambda^d_{ijk}$
are closely related to $\mu_0$ and $Y^d_{jk}$
by the value 
of  $l_i-h_1$.
There are then the following three possibilities.
\begin{table}
\caption{Bounds on the $\Lambda^d$ and $\Lambda^u$ from various
experiments. $\tilde{m}$ stands for typical squark or slepton mass.
}\medskip
\begin{center}
\begin{tabular}{|c|c|c|}\hline
 couplings & upper bound & experiment \\ \hline
 $\Lambda^d_{i12} \Lambda^d_{i21}$ 
  & $10^{-10}\, (\tilde{m}/\TeV)^2 \sim \lambda^{15}$ 
  & $\epsilon$ \cite{bm}  \\ \hline
 $\Lambda^d_{i13} \Lambda^d_{i31}$ 
  & $3.6\times10^{-7}\, (\tilde{m}/\TeV)^2 \sim \lambda^{10}$ 
  & $\delta m_B$ \cite{cr}  \\ \hline
 $\Lambda^u_{i13} \Lambda^u_{i23}$ for $i=2,3$  
  & $3\times10^{-4}\, (\tilde{m}/\TeV) \sim \lambda^5 $
  & $\epsilon$ \cite{bm} \\ \hline
  $\Lambda^d_{j12} \Lambda^e_{1jk}$ for $j=2,3$, $k=1,2$
  & $8\times 10^{-7} (\tilde{m}/\TeV)^2 \sim \lambda^9 $
  & $K_L \to e\bar{e}, \mu\bar{\mu}$ \cite{cr} \\ \hline
  $\Lambda^u_{11k}\Lambda^d_{ijk}$ for $j=1,2$
  & $10^{-24}\,(\tilde{m}/\TeV)^2 \sim \lambda^{37} $
  & proton decay  \\ \hline
 $\Gamma_{112}^i $
  & $10^{-8}(\tilde{m}/\TeV) \sim \lambda^{11} $
  & proton decay \cite{bn} \\ \hline
\end{tabular}
\end{center}
\end{table} 

(i) $l_i-h_1$ is a fractional number with
$N\geq 2$ in the denominator. In this case, the U(1)$_X$ charges
of the operators $L_iQ_jD^c_k$ and  $L_iH_2$ are fractional also, and thus 
neither $\Lambda_{ijk}^d$ nor $\mu_i$ 
are allowed. In order to see this, let $y^d_{ijk}$, $y^d_{ij}$, and $y_i$
denote the U(1)$_X$ charges of $L_iQ_jD^c_k$, $H_1Q_iD^c_j$, and
$L_iH_2$ respectively. Then 
\bea
&&y^d_{ijk}= l_i+q_j+d_k=(l_i-h_1)+ y^d_{jk}\,,
\nonumber \\
&&y_i= l_i+h_2=(l_i-h_1)-1\,.
\eea
Since $y^d_{jk}=
(q_{j3}+d_{k3})+x$ are all integers [see eq.~\refs{qch}],
obviously $y^d_{ijk}$ and $y_i$ are all fractional for  a fractional value
of $l_i-h_1$.

(ii)  $l_i-h_1$ is a negative integer or zero. In this case,
$\mu_i/\mu_0 \simeq \lambda^{|l_i-h_1|} \leq  1$.  
Let us first consider  the pattern (I).
If $y^d_{i11}$ is a non-negative integer and $l_i-h_1$ is not zero, 
the SUGRA superpotential
would give $\Lambda^d_{i11}\geq
\lambda^{x+3}$ since $0\leq y^d_{i11}\leq x+3$
from eq.~\refs{qch}.
Diagonalization of the K\"{a}hler metric then gives rise to 
$\Lambda^d_{ijk} \to \lambda^{|q_{j1}|+|d_{k1}|}
\Lambda^d_{i11}$, which leads to
$\Lambda^d_{i12}\simeq \Lambda^d_{i21}\geq \lambda^{x+4}$.
Obviously this is {\it inconsistent} with
the first experimental bound in   Table 1 for $0\leq x\leq 3$.
The same is trivially  true when $l_i-h_1=0$ for which 
$\Lambda^d_{i12}\simeq \Lambda^d_{i21}\simeq \lambda^{x+3}$.
We thus conclude that for the pattern (I),
$y^d_{i11}$ (and thus all charges $y^d_{ijk}$) should be negative,
and thus there is no contribution to $\Lambda^d_{ijk}$ 
from the SUGRA superpotential.
Since $y^d_{i11}=l_i-h_1+x+4$ from eq.~\refs{qch}, a negative integer
value of $y^d_{i11}$ implies 
\be
|l_i-h_1| \geq x+5.
\ee
In this case,  another contribution to
$\Lambda^d_{ijk}$ from the diagonalization of the K\"{a}hler
metric,  i.e. 
$\lambda^{|l_i-h_1|} Y^d_{jk}$, can  satisfy the bounds in the Table 1.
More explicitly, in this case,  we have 
\be \label{ldI}
 \mu_i/\mu_0\;\simeq\;\Lambda^d_{ijk}/Y^d_{jk} 
     \;\simeq\; \lambda^{|l_i-h_1|} 
  \;\leq\; \lambda^{5+x}  \,.
\ee
In the case of generic soft terms, the neutrino mass of order
$\mu_i^2/\mu_0$ is generated at tree-level \cite{hemp}--\cite{nar}.  
Therefore in the scheme under
consideration, we have
$m_\nu \simeq \lambda^{2|l_i-h_1|}\mu_0 \leq  30 \, \lambda^{2x} $ keV.  
If we assume the universality of soft-terms which may be
necessary to suppress the flavor changing neutral currents, there will
be a loop-suppression factor of order $10^{-5}\sim 10^{-6}$
\cite{cw,np},  so that $m_\nu \leq 0.1 \, \lambda^{2x} $ eV.

For the pattern (II), more possibilities are  allowed.
For $|l_i-h_1| \geq 10+x$, we get $ \mu_i/\mu_0$, $\Lambda^d_{ijk}/Y^d_{jk} 
\simeq \lambda^{|l_i-h_1|}$ as in the case of the pattern (I).
The resultant tree-level neutrino masses are $m_\nu \leq 7\times
10^{-3}\lambda^{2x}$ eV.
In addition to this, the cases with $l_i-h_1=0$ (only for $x=2$) and 
$|l_i-h_1|=5+x$  also fulfill the first and second bounds on $\Lambda^d$ 
in Table 1.
The  case with $l_i-h_1=0$ would be disfavored since it leads to
a too large neutrino mass
when soft terms are generic.
Independently of this point,
it turns out that the cases with the pattern (II) can 
{\it not} be compatible with 
the proton stability bound  for the range of U(1)$_X$ charges
not exceeding 15.

(iii) $l_i-h_1$ is a positive integer large enough to make
$\mu_i\simeq \lambda^{l_i-h_1-1}M_P\leq \mu_0$.
For this, we would need at least $l_i-h_1\geq 25$ if we take 
$\mu_0/M_P\simeq \lambda^{24}$. (Throughout this paper, we
will set $m_{3/2}/M_P\simeq \lambda^{23}$ and thus
$\mu_0/M_P\simeq \lambda m_{3/2}/M_P\simeq \lambda^{24}$.)
Later, we will see whether this case can be realized
for $l_i$ and $h_1$ whose magnitudes
are allowed to be as large as  15.
Before rotating away $\mu_iL_iH_2$ from the effective
superpotential, $\Lambda^d_{ijk}$ appears to be extremely
small since 
$y^d_{ijk} \geq x+25$ for $l_i-h_1\geq 25$.
However after rotating away the $\mu_i$ 
terms, we have
$ \Lambda^d_{ijk} \simeq Y^d_{jk} \mu_i/\mu_0
\simeq \lambda^{l_i-h_1-25} Y^d_{jk}$.
Then the first experimental bound of Table~1  demands for the pattern
(I) to satisfy $l_i-h_1-25 \geq 5-x$.  In the same way, 
for the pattern (II), we need
$l_i-h_1-25 \geq 2-x$. In summary,  we find
for the case of $l_i-h_1\geq 25$
\be \label{ldii}
 \mu_i/\mu_0 \;\simeq\; \Lambda^d_{ijk}/Y^d_{jk} 
     \;\simeq\; \lambda^{l_i-h_1-25} 
 \;\leq\; \left\{\ba{c} \lambda^{5-x} \qquad\mbox{(I)}\\
                  \lambda^{2-x} \qquad\mbox{(II)}\ea\right.
\ee
This case would allow larger $\mu_i$ (or $\Lambda^d_{ijk}$) than the case 
(ii) and thus larger neutrino masses.
However again we do not find any example of this class 
for the range of U(1)$_X$ charges not exceeding 15.

\bigskip

Similarly to the L violating couplings $\Lambda^d_{ijk}$, 
the B violating couplings $\Lambda^u_{ijk}$ 
are determined also by one parameter, $b_0 \equiv u_3+2d_3$.
For the pattern (I) and (II), the U(1)$_X$
charges $y^u_{ijk}$ of $U^c_i D^c_j D^c_k$  are given by 
\be \label{lu}
 {\rm (I)} \qquad
 b_0 + \left( \ba{ccc} 6 & 6 & 5 \\ 3 & 3 & 2 \\ 1 & 1 & 0 \ea \right)
   \,, \qquad {\rm (II)}\qquad
 b_0 + \left( \ba{ccc} 18 & 18 & 11 \\ 9 & 9 & 2 \\ 7 & 7 & 0 \ea \right)
\ee
where the low is for $(u_1,u_2,u_3)$ and the column is for
$(d_1 d_2, d_1 d_3, d_2 d_3)$.  Obviously, if $b_0$ is fractional,
$y^{u}_{ijk}$ are all fractional and thus $\Lambda^u_{ijk}$ are all
vanishing due to the unbroken $Z_N$. 
For the case that $b_0$ is an integer,
$K$-$\bar{K}$ mixing, i.e. the third bound in Table 1, 
sets a constraint on $b_0$ but 
{\it only} for the pattern  (I) as 
\be
b_0 \leq -4 \quad {\rm or}\quad  b_0 \geq 2.
\ee 
Here we do not consider the bound on $\Lambda^u_{ijk}$ coming from
the $n$--$\bar{n}$ 
oscillation \cite{zwir} or double nucleon decay \cite{bm} since 
it can be as large as order one for $\tilde{m} \simeq 1$ TeV 
and for a generous value of the hadronic scale \cite{gsher}.

\bigskip

When both $l_i-h_1$ and $b_0$ are integers,  the sum of them is 
constrained by proton stability.
For an integer $b_0$, $\Lambda^u_{11k} \simeq \lambda^{y^u_{11k}}$
for $y^u_{11k}\geq 0$ and
$\Lambda^u_{11k} \simeq \lambda^{-y^u_{11k}}m_{3/2}/M_P
\simeq  \lambda^{25-y^u_{11k}}$ for
$y^u_{11k}<0$, where
$y^u_{11k}=b_0+6$ and $b_0+18$ for (I) and (II) respectively.
Also $\Lambda^d_{i2k}\simeq \lambda^{x+2+n_i}$
where  the non-negative integer $n_i=h_1-l_i$ or $l_i-h_1-25$ 
from eqs.~\refs{ldI} or \refs{ldii}.
Then the proton stability condition reads
$n_i + y^u_{11k} \geq 35-x$ for $y^u_{11k} \geq 0$ 
and $n_i-y^u_{11k}+25 \geq 35-x$ for $y^u_{11k}<0$.

In addition to this, one also  has to consider the nonrenormalizable terms
in the effective superpotential (1).
For instance,  $\Gamma_{ijk}^l \simeq \lambda^{y^l_{ijk}}$
should be suppressed appropriately as in Table 1 when  
$y^l_{ijk} \equiv q_i +q_j+q_k+l_l$ is a nonnegative integer.
Rewriting $y^l_{ijk} = (l_i-h_1)-b_0+ 2x
+1+(q_{i3}+q_{j3}+q_{k3})$, one can see that the operators $Q_iQ_jQ_kL_l$
are allowed even when both $l_i-h_1$ and $b_0$ are fractional numbers
as long as their sum is an integer.
Combined with $\Lambda^d_{ijk}$,  the next
$d=5$ coupling $\Gamma^0_{ijk}$ can also 
induce a too fast proton decay unless 
$\Gamma^0_{12j}\Lambda^d_{ijk} \leq 10^{-8} \simeq \lambda^{11}$ 
for $k=1,2$.  
It turns out that this bound is simply satisfied as 
$\Gamma^0_{ijk} \leq \lambda^{11}$ in all the models which
pass  the proton stability conditions in Table 1.
About the third $d=5$ coupling $\Gamma^{'l}_{ijk}$, proton stability
bound depends upon the unknown  
mixing in the gluino couplings as
$\Gamma^{'l}_{1jk}(K^u_{RR})_{1j} \leq 10^{-9} \simeq \lambda^{12}$ 
for $j=2,3$ and $k,l=1,2$ 
\cite{bn}.
As we do not have any information on $K^u_{RR}$ in our approach, 
this bound will not be taken into account.
Combined with $\Lambda^u_{ijk}$, 
the coefficients  of other higher dimensional
operators like $[QU^c E^c H_1]_F$, $[QU^c \overline{L}]_D$ and 
$[U^c \overline{D^c} E^c]_D$
are  restricted also by the proton stability.
However in our scheme, typically
$\Lambda^u_{ijk}$ comes  from the K\"ahler term. 
As a result,  $\Lambda^u \leq m_{3/2}/M_P$ and
thus those bounds are
trivially satisfied.  

\bigskip

\section*{\ns\bf 4. Models}

Let us now  find some  
models, i.e. some U(1)$_X$ charge assignments,
satisfying all the bounds on B/L violating 
couplings in Table 1.
As seen from many superstring models \cite{charges}, we expect the 
U(1) charges are not too large.  
In the former investigations \cite{blr,cl},  
some examples satisfying the phenomenological bounds on
B/L violations are worked out, however  all of
them have ridiculously large U(1) charges.
If the maximum charge is limited  
to be less than 10 (for the smallest charge normalized to the unity),
we  find only one acceptable charge assignment (Model 1 of Table 2) 
with the pattern (I) and $N=1$.
In the last three columns of the Table, we provide 
the predicted size of the couplings which are relevant for  
the proton decay. One can see that, in Model 1,
proton can  decay with a rate not far below
the current experimental limit. This
is essentially due to the nonrenormalizable couplings  $\Gamma_{112}^{1,2}$
since  the renormalizable couplings
$\Lambda^u\Lambda^d$ are far below the current limit.
In fact, it is a generic feature of our scheme that
the nonrenormalizable couplings are somewhat close
to the current experimental limits.
In  Model 1, we have $b_0=u_3+2d_3=-12$ and thus
all the charges of $U^c_iD^c_jD^c_k$ are negative [see eq.~\refs{lu}]. 
As a result, $\Lambda^u_{ijk}$ arise only
from the SUGRA K\"{a}hler potential and thus are suppressed by
the extremely small factor $m_{3/2}/M_P\simeq \lambda^{23}$.
However the couplings $\Lambda^{d,e}_{ijk}$ can be induced 
through rotating away the off-diagonal K\"{a}hler metric
components $K_{L_iH_1^{\dagger}}$, 
yielding
$\Lambda^{d,e}_{ijk}\simeq \lambda^{|l_i-h_1|}Y^{d,e}_{jk}$.
\renewcommand{\arraystretch}{0.5}
\begin{table}
\caption{{\it Model 1.} U(1)$_X$ charges of the MSSM fields
in the range of maximum charge 10.  Here $N=1$, $x=3$. 
}\medskip
\begin{center}
\begin{tabular}{|c||c|c|c|c|c|c|c||c|c|c|}\hline
    &   &   &    &   &   &   &  &  &  & \\
  $i$ & $q_i$ & $u_i$ & $d_i$ & $l_i$ & $e_i$ & $h_1$ & $h_2$ 
  & $\Lambda^u_{11k}\Lambda^d_{32k}$ 
  & $\Gamma^{1,2}_{112}$ & $\Gamma^{'2}_{132}$ \\ 
    &   &   &    &    &   &  &  &  &  & \\ \hline
    &   &   &    &    &   &  &  &  &  & \\
  1 & 7 & 9 & -7 & -8 & 9 &  &  &  &  & \\
    &   &   &    &     &  &  &  &  &  & \\ \cline{1-6}
    &   &   &    &     &  &  &  &  &  & \\
  2 & 6 & 6 & -8 & -8 & 6 & 7& -8  & $\lambda^{45}$ 
                            & $\lambda^{12}$ & $\lambda^{11}$ \\  
    &   &   &    &    &   &  &  &  &  & \\ \cline{1-6}
    &   &   &    &    &   &  &  &  &  & \\ 
  3 & 4 & 4 & -8 & -4 & 0 &  &  &  &  & \\  
    &   &   &    &   &   &   &  &  &  & \\ \hline
\end{tabular}
\end{center}
\end{table}
\renewcommand{\arraystretch}{1.3}
In summary, the charged lepton Yukawa couplings  
in Model 1 are found to be
\be \label{Ye1}
Y^e \simeq \lambda^3 \left(\ba{ccc} 
                     \lambda^{5} & \lambda^{2} & \lambda^{4} \\
                     \lambda^{5} & \lambda^{2} & \lambda^{4} \\
                     \lambda^{9} & \lambda^{6} & 1 \ea\right),
\ee
and the magnitudes of the nonzero B/L violating couplings in Model 1
are given by 
\bea \label{model1} 
&& \Lambda^u_{ijk} \simeq \lambda^{29} \sim \lambda^{35},\quad 
 \Lambda^{d,e}_{1jk,2jk,3jk} \simeq 
        \lambda^{15,15,11}\, Y^{d,e}_{jk},
\quad \mu_{1,2,3}\simeq \lambda^{15,15,11} \mu_0, \nonumber\\
&& \Gamma^l_{ijk} \simeq \lambda^6 \sim  \lambda^{16},\quad 
 \Gamma^0_{ijk} \simeq \lambda^{21} \sim \lambda^{27},\quad 
 \Gamma^{'l}_{ijk} \simeq \lambda^{2} \sim \lambda^{17}. 
\eea
Since $\mu_i$ are very small, $\mu_i \leq \lambda^{11} \mu_0\simeq 6$ keV,
the tree-level neutrino masses  due to  $\mu_i$ 
are  negligible,  $m_{\nu}\leq  3\times 10^{-4} \eV$, 
even for generic  soft supersymmetry breaking  terms.  

When the maximum value of the U(1)$_X$  charges is relaxed up to 15, 
there appear basically two types of additional models and all of them 
possess an unbroken $Z_2$ subgroup of U(1)$_X$.
The first type of models allows $\Lambda^u_{ijk}$ to be nonvanishing
since 
the operators  $U^c_i D^c_j D^c_k$  are  $Z_2$ even,
while  the second type does not.
In Table 3, we show a representative model (Model 2) of the first 
type which 
has  the following lepton Yukawa couplings
\be \label{Ye2}
Y^e \simeq 
  \left(\ba{ccc} \lambda^{5} & \lambda^5 & 0 \\
                 \lambda^2 & \lambda^{2} & 0 \\
                 0 & 0 & 1 \ea\right),
\ee
and the nonvanishing B/L violating couplings:
\bea \label{model2}
&&\Lambda^u_{ijk} \simeq \lambda^{27} \sim \lambda^{33}\,, \quad 
 \Lambda^{d}_{3jk} \simeq \lambda^{8}\, Y^{d}_{jk}\,,\quad
 \Lambda^{e}_{131,\,132,\,231,\,232}  \simeq \lambda^{13,\,13,\,10,\,10}, 
         \nonumber \\
&& \mu_3\simeq \lambda^8\,\mu_0 \,,\quad
\Gamma^3_{ijk} \simeq \lambda^5 \sim  \lambda^{11},\quad 
 \Gamma^{0}_{ijk} \simeq \lambda^{13} \sim  \lambda^{19},\quad 
 \Gamma^{'3}_{ijk} \simeq \lambda \sim \lambda^{7}  \,.
\eea
Notice that $U^c$, $D^c$, $L_{1,2}$, $E^c_3$ 
are $Z_2$ even, while the others are odd. Therefore after
U(1)$_X$ breaking into $Z_2$, the operators like $L_3 Q D^c$, 
$U^c D^c D^c$, and $QQQL_3$ can be induced, leading to 
the proton decay.
\renewcommand{\arraystretch}{0.5}
\begin{table}
\caption{{\it Model 2.} U(1)$_X$ charges of the MSSM fields 
in the range of maximum charge 15.  Here $N=2$ and $x=0$. 
}\medskip
\begin{center}
\begin{tabular}{|c||c|c|c|c|c|c|c||c|c|c|}\hline
    &   &   &    &   &   &   &  &  &  & \\
  $i$ & $q_i$ & $u_i$ & $d_i$ & $l_i$ & $e_i$ & $h_1$ & $h_2$ 
  & $\Lambda^u_{11k}\Lambda^d_{32k}$
  & $\Gamma^{3}_{112}$ & $\Gamma^{'1,2}_{132}$ \\ 
    &   &   &    &    &   &  &  &  &  & \\ \hline
    &   &   &    &    &   &  &  &  &  & \\
  1 & 11/2 & 7 & -5 & -4 & 11/2 &  &  &  &  & \\
    &   &   &    &     &  &  &  &  &  & \\ \cline{1-6}
    &   &   &    &     &  &  &  &  &  & \\
  2 & 9/2 & 4 & -6 & -7 & 11/2 & 7/2 & -9/2  
              & $\lambda^{37}$ & $\lambda^{11}$ & $ 0 $  \\  
    &   &   &    &    &   &  &  &  &  & \\ \cline{1-6}
    &   &   &    &    &   &  &  &  &  & \\ 
  3 & 5/2 & 2 & -6 & -9/2 & 1 &  &  &  & & \\  
    &   &   &    &   &   &   &  &  &  & \\ \hline
\end{tabular}
\end{center}
\end{table}
\renewcommand{\arraystretch}{1.3}
Contrary to Model 1,  proton life-time is on the verge of the 
experimental limit due to the larger values of 
both $\Lambda^u_{11k} \Lambda^d_{32k}$ and $\Gamma^3_{112}$.
Neutrino mass in Model 2 can be large as
$m_{\nu_1} \simeq \lambda^{12}\mu_0 \simeq 3 \keV$ for generic
soft terms.
It could in fact be smaller than the cosmological bound 
($m_\nu \ler 100$ eV) depending upon other parameters of the theory
\cite{hemp}--\cite{nar}.  In addition to Model 2,
there is another model of first type  with almost same properties 
except that e.g.
$\mu_2$ is allowed instead of $\mu_3$.

The models of the second type have $Z_2$ parity which forbids both 
$U^c D^c D^c$ and $QQQH_1$.  In particular, the representative model 
(Model 3) in Table 4 has  the $Z_2$ parity under which
$Q$, $L_2$, $E^c_{1,3}$ are 
even and the other are odd.
In Model 3, the charged lepton Yukawa couplings are given by
\be \label{Ye3}
Y^e \simeq 
  \lambda^3 \left(\ba{ccc} 
                     \lambda^{5} & 0 & 1 \\
                     0 & \lambda^{2} & 0 \\
                     \lambda^{5} & 0 & 1\ea\right) \,.
\ee
and 
the magnitudes  of the nonvanishing  B/L violating couplings are 
\bea \label{model3}
&&\Lambda^{d}_{1jk,3jk} \simeq \lambda^{8}\,Y^{d}_{jk}\,,\quad
 \Lambda^{e}_{131,\,122,\,232,\,133}  \simeq \lambda^{0,\,13,\,13,\,5}, 
\nonumber \\
&& \mu_{1,3}\simeq \lambda^8 \mu_0 \,,
\quad \Gamma^2_{ijk} \simeq \lambda^7 \sim  \lambda^{13},\quad 
 \Gamma^{'2}_{ijk} \simeq \lambda^7 \sim \lambda^{13}  \,.
\eea
The neutrino mass from $\mu_{1,3}\simeq \lambda^{8} \mu_0$ is  
$m_{\nu} \simeq 3 \eV$.  As mentioned earlier, the neutrino masses 
are further suppressed if
soft terms satisfy certain universality conditions.
We found also three more models of the second type in which
$\Gamma^l_{112}$ are in the range of $\lambda^{15}\sim\lambda^{13}$ and 
$\Gamma'^l_{ijk}$ are usually much larger than $\lambda^{10}$.
Therefore, the proton decay rate in the second type models
is smaller than the previous models by 
factor of $\lambda^{2,4}$ or less.
\renewcommand{\arraystretch}{0.5}
\begin{table}
\caption{{\it Model 3.} U(1)$_X$ charges of the MSSM fields 
in the range of maximum charge 15.  Here $N=2$ and $x=3$.
}\medskip
\begin{center}
\begin{tabular}{|c||c|c|c|c|c|c|c||c|c|c|}\hline
    &   &   &    &   &   &   &  &  &  & \\
  $i$ & $q_i$ & $u_i$ & $d_i$ & $l_i$ & $e_i$ & $h_1$ & $h_2$ 
  & $\Lambda^u \Lambda^d $
  & $\Gamma^2_{112}$ & $\Gamma^{'2}_{132}$ \\ 
    &   &   &    &    &   &  &  &  &  & \\ \hline
    &   &   &    &    &   &  &  &  &  & \\
  1 & 6 & 15/2 & -7/2 & -7/2 & 7 &  &  &  &  & \\
    &   &   &    &     &  &  &  &  &  & \\ \cline{1-6}
    &   &   &    &     &  &  &  &  &  & \\
  2 & 5 & 9/2 & -9/2 & -4 & 9/2 & 9/2 & -11/2   
    & 0 &   $\lambda^{13}$ & $\lambda^{10}$ \\  
    &   &   &    &    &   &  &  &  &  & \\ \cline{1-6}
    &   &   &    &    &   &  &  &  &  & \\ 
  3 & 3 & 5/2 & -9/2 & -7/2 & 2 &  &  &  & & \\  
    &   &   &    &   &   &   &  &  &  & \\ \hline
\end{tabular}
\end{center}
\end{table}
\renewcommand{\arraystretch}{1.3}

Let us finally comment on the couplings $\Gamma'^l_{ijk}$.
The bound on this coupling from the 
proton decay depends upon
the mixing in the gaugino couplings: 
$\Gamma'^l_{1jk}(K^u_{RR})_{1j}\leq \lambda^{12}$
for $j=2,3$ and $k,l=1,2$.
Models 1 and 3 can be consistent with this constraint 
when the flavor mixing in gaugino couplings are 
$(K^u_{RR})_{1j} \simeq \lambda^{1,2}$ respectively,
while in Model 2 even an arbitrary flavor mixing would not cause proton decay.
If it is required, one could make the flavor mixing 
small in our scheme for instance
by assuming the usual universality of
soft terms at $M_P$.

\bigskip

\section*{\ns\bf 5. Conclusion}

To summarize, we suggest the relevance of an anomalous horizontal
abelian symmetry for the resolution of all the hierarchy problems
in the supersymmetric standard model, viz
the quark and lepton mass hierarchy, the $\mu$ problem,
and the highly suppressed 
(both renormalizable and nonrenormalizable)
B/L violating  interactions. 
This anomalous U(1)$_X$ would be
a gauge symmetry as found in many superstring models
endowed with the Green-Schwarz anomaly cancellation mechanism.
In view of various string model constructions, the magnitudes of 
U(1)$_X$  charges are not likely to be so large.  
Observed quark masses and
mixings, lepton masses and several experimental bounds on B/L
violating couplings are used  together with the assumption that
$\mu_0\simeq \lambda m_{3/2}$ in order 
to single out only a few models with
reasonable charge assignments.  For the most acceptable charge assignment 
allowing the biggest U(1)$_X$ charge to be 9,
only one model with $N=1$ (Model 1)
is found.   In this model, renormalizable B/L violating  
couplings (including $\mu_i$) are
extremely suppressed, and thus not yield any observable  signature.  
On the other
hand, 
the coefficients $\Gamma^{1,2}_{112}\simeq \lambda^{12}$ of 
nonrenormalizable $d=5$ operators
$Q_iQ_jQ_kL_l$ are relatively large, so that
may render proton decay observable in the near future. 

Relaxing the limit of  U(1)$_X$ charges to 15, we found 
two types of additional  models with an unbroken $Z_2$ parity, 
i.e. models with
$N=2$.  
The models  of the first type of allow both $L_iQ_jD^c_k$ and 
$U^c_i D^c_j D^c_k$ after the U(1)$_X$ breaking into
$Z_2$. They  predict a marginally detectable proton decay 
due to the renormalizable
couplings  $\Lambda^u_{11k}\Lambda^d_{i2k}\simeq \lambda^{37}$  
as well as the nonrenormalizable couplings $\Gamma^{1,2}_{112} \simeq
\lambda^{11}$ which are on the verge of the current bound.
In the second type of models, 
$U^c_i D^c_j D^c_k$ are  $Z_2$ odd and thus  are 
completely forbidden. Its representative model (Model 3)
has $\Gamma^2_{112}\simeq \lambda^{13}$
which is away from the proton stability bound by the factor
of $\lambda^2$.
In our scheme, renormalizable B/L violating terms tend to 
be highly suppressed, while 
nonrenormalizable couplings are not far from the current
experimental limits.
In particular, the operators $U^c_i U^c_j D^c_k E^c_l$ may have 
coefficients
larger than $\lambda^{12}$, and then
(approximate) squark degeneracy has to be implemented for
the proton stability.
The $\mu_i$ are typically small enough to yields cosmologically 
safe neutrino masses even for generic forms of  soft supersymmetry
breaking.  We however find no models with  L violation patterns
providing solar or atmospheric neutrinos.

\bigskip

{\bf Acknowledgments}:
EJC thanks Y. Grossmann, H. P. Nilles and P. L. White 
for useful communications.
This work is supported in part by KOSEF Grant 951-0207-002-2 (KC, HK),
KOSEF through CTP of Seoul National University (KC),
Programs of Ministry of Education BSRI-96-2434 (KC), Non Directed Research
Fund 
of Korea Research Foundation, 1996 (EJC).  EJC is a Brain-Pool fellow.

\end{document}